\documentclass[10pt,a4paper,twoside]{article}

\usepackage{amssymb,amsmath,multirow,epstopdf,graphicx}

\newtheorem{theorem}{Theorem}[section]

\newtheorem{definition}[theorem]{Definition}

\newcommand{\bs}{\boldsymbol}

\begin{document}

\title{Comparison of Efficiencies of some Symmetry Tests around an Unknown Center}

\author{  Bojana Milo\v sevi\'c\footnote{bojana@matf.bg.ac.rs} , Marko Obradovi\' c \footnote{marcone@matf.bg.ac.rs}
         \\\medskip
{\small Faculty of Mathematics, University of Belgrade, Studenski trg 16, Belgrade, Serbia}}

\date{}

\maketitle




\begin{abstract}
	In this paper, some recent and classical tests of symmetry are modified for the case of an unknown center. 
	The unknown center is estimated with its $\alpha$-trimmed mean estimator.
	The asymptotic behavior of the new tests is explored. 
	The local approximate Bahadur efficiency is used to compare the tests to each other as well as to some other tests.
\end{abstract}

{\small \textbf{ keywords:} U-statistics with estimated parameters; $\alpha$-trimmed mean; asymptotic  efficiency

}



\section{Introduction}

The problem of testing symmetry has been popular for decades, mainly due to the fact
that many statistical methods depend on the assumption of symmetry.
The well-known  examples are the robust estimators of location such as trimmed means that implicitly assume
that the data come from a symmetric distribution. Another example are the bootstrap confidence intervals, that tend to converge faster when 
the corresponding pivotal quantity is
symmetrically distributed.

Probably the most  famous symmetry tests are the classical sign and Wilcoxon tests, as well as the tests proposed following the example of Kolmogorov-Smirnov
and Cramer-von Mises statistics (see. \cite{smirnov1947,butler1969test,rothman1972cramer}).
Other examples of papers on this topic include
\cite{maesono1987competitors, ahmad1997testing, burgioNikitin0, burgioNikitin, baklizi2007testing, burgio2011most, nikitinAhsanullah, milosevicObradovicSimetrijaSPL, amiri2016new}.
%Some  symmetry tests can be found in e.g. \cite{burgioNikitin0} and \cite{burgioNikitin}.

All of these symmetry tests are designed for the case of a known center of distribution. They share many nice properties such as distribution-freeness under the null
symmetry hypothesis.

There have been many attempts to adapt  these tests to the case of an unknown center.
Some modified Wilcoxon tests can be found  in \cite{bhattacharya1982two,antille1977tests,antille1982testing} and a modified sign test in 
\cite{gastwirth1971sign}.

There also exist some symmetry tests originally designed for testing symmetry around an unknown center.    The famous $\sqrt{b_1}$ test is one of the examples.
Some other tests have been proposed in \cite{mira1999distribution,cabilio1996simple, Gastwirth}.

The goal of our paper is to compare some symmetry tests around an unknown center. Instead of commonly used power comparison 
(see e.g. \cite{Gastwirth,zheng2010bootstrap,mira1999distribution,farrell2006comprehensive}),
here we compare the tests using the local asymptotic efficiency.
We opt for the approximate Bahadur efficiency since it is applicable to asymptotically non-normally distributed test statistics.
The Bahadur efficiency of symmetry tests has been considered in, among others, \cite{beghin1999approximate, meintanis2007testing,henze2009integral,milovsevic2016new}.

Consider the setting of testing the null hypothesis $H_0:\theta\in\Theta_0$ against the alternative
$H_1:\theta\in\Theta_1$. Let us suppose that for a test statistic $T_n$, under $H_0$, the limit 
$\lim_{n\rightarrow\infty}P\{T_n\leq t\}=F(t)$, where $F$ is non-degenerate distribution function, exists. 
Further, suppose that $\lim_{t\rightarrow\infty}t^{-2}\log(1-F(t))=-\frac{a_T}{2}$, and
that the limit in probability $P_\theta$ $\lim_{n\rightarrow\infty}T_n=b_T(\theta)>0$, exists for $\theta \in \Theta_1$. 
The relative approximate Bahadur efficiency with respect to another test statistic $V_n$ is
\begin{equation*}
e^{\ast}_{T,V}{(\theta)}=\frac{c^{\ast}_T{(\theta)}}{c^{\ast}_V{(\theta)}},
\end{equation*}
where \begin{equation}\label{appslope}c^{\ast}_T(\theta)=a_Tb_T^2(\theta)\end{equation} 
is the approximate Bahadur slope of $T_n$. Its limit when $\theta\to 0$ is called the local approximate Bahadur efficiency.

The tests we consider may be classified into two groups according to their limiting distributions:
asymptotically normal ones; and those whose asymptotic distribution coincides with the supremum of some Gaussian process.

For the first group of tests, the coefficient $a_T$ is the inverse of the limiting variance. For the second, it is the inverse of the supremum of the covariance
function of the limiting process (see \cite{marcus1972sample}).
\section{Test statistics}
\label{sec:1}

Most of considered tests are obtained by modifying the symmetry tests around known location parameter. 
Let $X_1,..,X_n$ be an i.i.d. sample with distribution function $F$. The tests are applied  to the sample shifted by the value
of the location estimator. Typical choices of location estimators are the mean and the median. Here we take a more general approach,
using $\alpha$-trimmed means

\begin{equation*}
\mu(\alpha)=\frac{1}{1-2\alpha}\int_{F^{-1}(\alpha)}^{F^{-1}(1-\alpha)}xdF(x),\;\;0<\alpha<1/2,
\end{equation*}
including their boundary cases $\mu(0)$ -- the mean and $\mu(1/2)$ --  the median. The estimator is

\begin{equation}\label{alfasamplemean}
\widehat{\mu}(\alpha)=\frac{1}{1-2\alpha}\int_{F_n^{-1}(\alpha)}^{F_n^{-1}(1-\alpha)}xdF_n(x),\;\;0<\alpha<1/2.
\end{equation}

In the case of $\alpha=0$ and $\alpha=1/2$, the estimators are the sample mean and the sample median, respectively.

The modified statistics we consider in this paper are:

\begin{itemize}
	\item Modified sign test
	\begin{equation*}
	{\rm S}=\frac{1}{n}\sum_{i=1}^n {\rm I}\{X_j-\widehat{\mu}(\alpha)>0\}-\frac12;
	\end{equation*}
	
	\item Modified Wilcoxon test
	\begin{equation*}
	{\rm W}=\frac{1}{\binom{n}{2}}\sum_{1\leq i<j\leq n} {\rm I}\{X_i+X_j-2\widehat{\mu}(\alpha)>0\}-\frac12;
	\end{equation*}
	
	\item Modified Kolmogorov-Smirnov symmetry test
	
	\begin{equation*}
	{\rm KS}=\sup_{t}|F_n(t+\widehat{\mu}(\alpha))+F_n(\widehat{\mu}(\alpha)-t)-1|;
	\end{equation*}
	
	\item Modified tests based on the Baringhaus-Henze characterization (see \cite{baringhausHenzeSymmetry,litvinovaSimetrija})
	
	\begin{align*}
	{\rm BH}^{I}&=\frac1{n\binom n2}\sum_{i_3=1}^n\sum_{\mathcal{C}_2}\Big(\frac12{\rm I}\{|X_{i_1}-\widehat{\mu}(\alpha)|<|X_{i_3}-\widehat{\mu}(\alpha)|\}+\frac12{\rm I}\{|X_{i_2}-\widehat{\mu}(\alpha)|<|X_{i_3}-\widehat{\mu}(\alpha)|\}\\&-
	{\rm I}\{|X_{2;X_{i_1},X_{i_2}}-\widehat{\mu}(\alpha)|<|X_{i_{3}}-\widehat{\mu}(\alpha)|\}\Big);\\
	{\rm BH}^{K}&=\sup_{t>0}\Big|\frac1{\binom n2}\sum_{\mathcal{C}_2}\Big(\frac12{\rm I}\{|X_{i_1}-\widehat{\mu}(\alpha)|<|X_{i_3}-\widehat{\mu}(\alpha)|\}+\frac12{\rm I}\{|X_{i_2}-\widehat{\mu}(\alpha)|<|X_{i_3}-\widehat{\mu}(\alpha)|\}\\&-
	{\rm I}\{|X_{2;X_{i_1},X_{i_2}}-\widehat{\mu}(\alpha)|<t\}\Big)\Big|;
	\end{align*}
	
	\item Modified tests based on the Ahsanullah's characterization (see \cite{nikitinAhsanullah})
	
	\begin{align*}
	{\rm NA}^{I}(k)&=\frac1{n\binom nk}\sum_{i_{k+1}=1}^n\sum_{\mathcal{C}_k}\Big({\rm I}\{|X_{1;X_{i_1},\ldots,X_{i_k}}-\widehat{\mu}(\alpha)|<|X_{i_{k+1}}-\widehat{\mu}(\alpha)|\}\\&-
	{\rm I}\{|X_{k;X_{i_1},\ldots,X_{i_k}}-\widehat{\mu}(\alpha)|<|X_{i_{k+1}}-\widehat{\mu}(\alpha)|\}\Big);\\
	{\rm NA}^{K}(k)&=\sup_{t>0}\Big|\frac1{\binom nk}\sum_{\mathcal{C}_k}\Big({\rm I}\{|X_{1;X_{i_1},\ldots,X_{i_k}}-\widehat{\mu}(\alpha)|<t\}\\&-
	{\rm I}\{|X_{k;X_{i_1},\ldots,X_{i_k}}-\widehat{\mu}(\alpha)|<t\}\Big)\Big|;
	\end{align*}
	
	\item Modified tests based on the Milo\v sevi\' c-Obradovi\' c characterization (see \cite{milosevicObradovicSimetrijaSPL})
	
	\begin{align*}
	{\rm MO}^{I}(k)&=\frac1{n\binom n{2k}}\sum_{i_{2k+1}=1}^n\sum_{\mathcal{C}_{2k}}\Big({\rm I}\{|X_{k;X_{i_1},\ldots,X_{i_{2k}}}-\widehat{\mu}(\alpha)|<|X_{i_{2k+1}}-
	\widehat{\mu}(\alpha)|\}\\&-
	{\rm I}\{|X_{k+1;X_{i_1},\ldots,X_{i_{2k}}}-\widehat{\mu}(\alpha)|<|X_{i_{2k+1}}-\widehat{\mu}(\alpha)|\}\Big);\\
	{\rm MO}^{K}(k)&=\sup_{t>0}\Big|\frac1{\binom nk}\sum_{\mathcal{C}_{2k}}\Big({\rm I}\{|X_{k;X_{i_1},\ldots,X_{i_{2k}}}-\widehat{\mu}(\alpha)|<t\}\\&-
	{\rm I}\{|X_{k+1;X_{i_1},\ldots,X_{i_{2k}}}-\widehat{\mu}(\alpha)|<t\}\Big)\Big|,
	\end{align*}
	
\end{itemize}
where $X_{k;X_{i_1},...X_{i_m}}$ stands for the $k$th order statistic of the subsample $X_{i_1},...,X_{i_m}$, and $\mathcal{C}_{m}=\{(i_1,\ldots, i_m) : 1\leq i_1 < \cdots < i_m \leq n\}$.

Note that ${\rm NA}^{I}(2)$ and ${\rm MO}^{I}(1)$ coincide. The same holds for ${\rm NA}^{K}(2)$ and ${\rm MO}^{K}(1)$.

Among the tests originally intended for testing symmetry around an unknown mean, the most famous is  classical $\sqrt{b_1}$ test,  based on
the sample skewness coefficient, with test statistic
% $\sqrt{b_1}$ test proposed
%in \cite{}

\begin{equation}
\sqrt{b_1}=\frac{\hat{m}_3}{s^3},
\end{equation}
where $\hat{m}_3$ is the  sample  third central moment and $s$ is sample standard deviation.
The test is applicable if the sample comes from a distribution with  finite sixth moment.

We also consider the class of tests based on so-called Bonferoni measure. In \cite{cabilio1996simple} the following test statistic is proposed:

\begin{equation*}
{\rm CM}=\frac{\bar{X}-\hat{M}}{s},
\end{equation*}
where $\hat{M}$ is the sample median and $s$ is the sample standard deviation.
Similar tests are proposed in \cite{mira1999distribution} and \cite{Gastwirth}, with the following statistic 
\begin{align*}
&\gamma=2(\bar{X}-\hat{M})\\  &{\rm MGG}=\frac{\bar{X}-\hat{M}}{J},\;\;\text{where }
J=\sqrt{\frac{\pi}{2}}\frac{1}{n}\sum_{i=1}^n|X_i-\hat{M}|.
\end{align*}	
These tests are applicable if the sample comes from a distribution with finite second moment.

For the supremum-type tests, we consider large values of test statistic to be significant. For others, which are asymptotically normally distributed, 
we consider large absolute values of tests statistic to be significant.

\section{Bahadur approximate Slopes}

We can divide the considered test statistics into three groups based on their structure: 
\begin{itemize}
	\item non-degenerate U-statistics with estimated parameters;
	\item the suprema of families of non-degenerate (in sense of \cite{nikitinLDKS}) U-statistics  with estimated parameters;
	\item other statistics with limiting normal distribution.
\end{itemize}

Since we are dealing with U-statistics with estimated location parameters, we shall examine their limiting distribution using the technique 
from \cite{randles1982asymptotic}. With this in mind, we give the following definition.

\begin{definition}
	Let $\mathcal{U}$ be the family of U-statistics $U_n(\mu)$, with bounded symmetric kernel $\Phi(\cdot;\mu)$, that satisfy the following conditions:
	\begin{itemize}
		\item $EU_n(\mu)=0$;
		\item $\sqrt{n}U_n$ converges to normal distribution whose variance does not depend on $\mu$;
		\item For $K(\mu,d)$, a neighbourhood of $\mu$ of a radius $d$, there exists a constant $C>0$, such that,
		if $\mu'\in K(\mu,d)$ then $$E\big(\sup_{\mu'\in K(\mu,d)}|\Phi(\cdot;\mu')-\Phi(\cdot;\mu)|\big)\leq Cd.$$  \end{itemize}
\end{definition}

All our U-statistics belong to this family due to unbiasedness, non-degeneracy and uniform boundness  of their kernels.

Since our comparison tool is the local asymptotic efficiency, we are going to consider alternatives close to some symmetric distribution. 
Therefore, we define the family of close alternatives that satisfy some regularity conditions (see also \cite{nikitinMetron}).

\begin{definition}
	Let $\mathcal{G}=\{G(x;\theta)\}$ be the family of absolutely continuous distribution functions, with densities $g(x;\theta)$, 
	satisfying the following conditions:
	\begin{itemize}
		\item $g(x;\theta)$ is symmetric around some location parameter  $\mu$   if and only if $\theta=0$;
		\item $g(x;\theta)$ is twice continuously differentiable along $\theta$ in some neighbourhood of zero;
		\item all second derivatives of $g(x;\theta)$  exist  and are absolutely integrable for $\theta$ in some neighbourhood of zero.
	\end{itemize}
\end{definition}

For brevity, in what follows, we shall use the following notation: 
\begin{align*}
F(x):=G(x,0);\;\;f(x):=g(x,0);\;\;H(x):=G'_{\theta}(x,0);\;\;h(x):=g'_{\theta}(x,0).
\end{align*}  

The null hypothesis of symmetry can now be expressed as: $H_0:\theta=0.$
To calculate the local approximate slope \eqref{appslope}, we need to find the variance of the limiting normal distribution under the null hypothesis, 
as well as the limit in probability under a close alternative. We achieve this goal using  the following two theorems.

\begin{theorem}\label{disperzijaModifikovane} Let $\bs{X}=(X_1,X_2...,X_n)$ be an i.i.d. sample from an absolutely continuous symmetric distribution, 
	with distribution function $F$.
	Let $U_n(\mu)$ with kernel $\Phi(\bs{X};\mu)$ be a $U$-statistic of order $m$ from the family $\mathcal{U}$; and let $\widehat{\mu}(\alpha)$, $0\leq \alpha\leq 1/2$, 
	be the $\alpha$-trimmed sample mean \eqref{alfasamplemean}.
	Then $\sqrt{n}U_n(\widehat{\mu}(\alpha))$ converges in distribution to a zero mean normal random variable with the following variance:
	\begin{align*}
	\sigma^2_{U,F}(\alpha)&=m^2\Bigg(\int_{-\infty}^{\infty}\varphi^2(x)f(x)dx+\frac{2}{(1-2\alpha)^2}\bigg(\int_{-\infty}^{\infty}\varphi(x)f'(x)dx\bigg)^2\\&
	\cdot\bigg(\int_{0}^{q_{1-\alpha}}x^2f(x)dx+\alpha(q_{1-\alpha})^2\bigg)
	+\frac{4}{1-2\alpha}\bigg(\int_{-\infty}^{\infty}\varphi(x)f'(x)dx\bigg)\\&\cdot\bigg(\int_{0}^{q_{1-\alpha}}\varphi(x)xf(x)dx+
	q_{1-\alpha}\int_{q_{1-\alpha}}^{\infty}\varphi(x)f(x)dx\bigg)\Bigg),
	\end{align*} for $0<\alpha<1/2$, where  $\varphi(x)=E(\Phi(\mathbb{X};\mu)|X_1=x)$ is the first projection of the kernel  $\Phi(\mathbb{X};\mu)$  on a basic observation, 
	and $q_{1-\alpha}=F^{-1}(1-\alpha)$ is the $(1-\alpha)$th quantile of F.
	
	In the case of boundary values of $\alpha$, the expression above becomes:
	
	\begin{align*}
	\sigma^2_{U,F}(0)&=m^2\Bigg(\int_{-\infty}^{\infty}\varphi^2(x)f(x)dx+
	2\bigg(\int_{-\infty}^{\infty}\varphi(x)f'(x)dx\bigg)^2\bigg(\int_{0}^{\infty}x^2f(x)dx\bigg)
	\\&+4\bigg(\int_{-\infty}^{\infty}\varphi(x)f'(x)dx\bigg)\cdot\bigg(\int_{0}^{\infty}\varphi(x)xf(x)dx
	\bigg)\Bigg),
	\end{align*}
	and
	\begin{align*}
	\sigma^2_{U,F}(1/2)&=m^2\Bigg(\int_{-\infty}^{\infty}\varphi^2(x)f(x)dx+\frac{1}{4f^2(0)}
	\bigg(\int_{-\infty}^{\infty}\varphi(x)f'(x)dx\bigg)^2
	\\&+\frac{2}{f(0)}\bigg(\int_{-\infty}^{\infty}\varphi(x)f'(x)dx\bigg)\cdot\bigg(\int_{0}^{\infty}\varphi(x)f(x)dx\bigg)\Bigg).
	\end{align*}
	%Then
	%$\sqrt{n}(U_n(\mu),\widehat{\mu}(\alpha)-\mu)$ converges in distribution to bivariate normal $\mathcal{N}({\bf 0},\Sigma)$, where

\end{theorem}
\textbf{Proof.} We prove only the case $0<\alpha<0.5$. The rest are analogous and simpler.   

Notice that $\hat{\mu}(\alpha)$ has its Bahadur representation \cite{serfling}

\begin{equation}\label{bahadurRepresentation}
 \hat{\mu}(\alpha)-\mu(\alpha)=\frac{1}{n}\sum_{j=1}^n\psi_{\alpha,F}(X_j)+R_n,
\end{equation}
where
\begin{equation*}
\psi_{\alpha,F}(x)=\frac{1}{1-2\alpha}\int_{\alpha}^{1-\alpha}\frac{t-I\{x<F^{-1}(t)\}}{f(F^{-1})(t)}dt
\end{equation*}
is the influence curve of $\mu(\alpha)$, and $\sqrt{n}R_n$ converges in probability to zero.

Using the multivariate central limit theorem for U-statistics we conclude that the joint limiting  distribution of  $\sqrt{n}U_n$ and   
$\sqrt{n}(\hat{\mu}(\alpha)-\mu(\alpha))$ is bivariate normal $\mathcal{N}_2(\bs{0},\Sigma)$,
where
\begin{equation*}
\Sigma=\left(\begin{array}{cc}
m^2\int_{-\infty}^{\infty}\varphi^2(x)dF(x)&m\int_{-\infty}^{\infty}\psi_{\alpha,F}(x)\varphi(x)dF(x)\\
m\int_{-\infty}^{\infty}\psi_{\alpha,F}(x)\varphi(x)dF(x)&\int_{-\infty}^{\infty}\psi^2_{\alpha,F}(x)dF(x)
\end{array}\right).
\end{equation*}

Therefore, the conditions 2.3 and 2.9B of \cite[Theorem 2.13]{randles1982asymptotic} are satisfied.
Hence we have $  \sqrt{n}U_n(\widehat{\mu}(\alpha))\overset{d}{\rightarrow}\mathcal{N}(0,\sigma^2_{U,F}(\alpha))$ where

\begin{align*}
\sigma^2_{U,F}(\alpha)=[1, A]^T\Sigma[1, A], \text{ and }
A=E_{\gamma}\Phi(\cdot;\mu)^{'}_{\mu}\Big |_{\mu=\gamma}=m\int_{-\infty}^{\infty}\varphi(x)f'(x)dx.
\end{align*}
\hfill{$\Box$}
\begin{theorem}\label{bocenjeno}  Under the same assumptions as in Theorem \ref{disperzijaModifikovane}, the limit in probability of 
	the modified statistic $U_n(\widehat{\mu}(\alpha))$
	under alternative $g(x;\theta)\in \mathcal{G}$ is
	
	\begin{align*}
	b(\theta,\alpha)=m\int_{-\infty}^{\infty}\varphi(x)(h(x)+\mu'_{\theta}(0,\alpha)f'(x))dx\cdot\theta+o(\theta),
	\end{align*}
	where $\mu'_{\theta}(0,\alpha)=\frac{1}{1-2\alpha}\bigg(-q_{1-\alpha}\Big(H(q_{1-\alpha})+H(-q_{1-\alpha})\Big)+
	\int_{-q_{1-\alpha}}^{q_{1-\alpha}}xh(x)dx\bigg)$, for $0<\alpha<1/2$, $\mu'_{\theta}(0,1/2)=-H(0)/f(0)$, and 
	$\mu'_{\theta}(0,0)=\int_{-\infty}^{\infty}xh(x)dx$.
\end{theorem}
\textbf{Proof.}Let $L(\bs{x};\theta)$ be the likelihood function of the sample.
Using the law of large numbers for $U$-statistics with estimated parameters (see \cite{iverson}), 
the limit in probability of $U_n(\widehat{\mu}(\alpha))$ is

\begin{align*}
b(\theta,\alpha)&=\int_{-\infty}^{\infty}\Phi(\bs{x}-\mu(\theta,\alpha))L(\bs{x};\theta)d\bs{x}\\
&=\int_{-\infty}^{\infty}\Phi(\bs{x})L(\bs{x}+\mu(\theta,\alpha);\theta)d\bs{x}\\
&=\int_{-\infty}^{\infty}\Phi(\bs{x})\prod_{i=1}^n g(x_i+\mu(\theta,\alpha);\theta)d\bs{x}.
\end{align*}

The first derivative with respect to $\theta$ at $\theta=0$ is

\begin{align*}
b'(0,\alpha)&=\int_{-\infty}^{\infty}\Phi(\bs{x})\Big[\sum_{j=1}^n(\mu'(0,\alpha)f'(x_j)+h(x_j))\prod_{\underset{i\neq j}{i=1}}^nf(x_i)\Big]d\bs{x}\\
&=m\int_{-\infty}^{\infty}\varphi(x)((\mu'(0,\alpha)f'(x)+h(x))dx.
\end{align*}
Expanding $b(\theta,\alpha)$ in the Maclaurin series we complete the proof. \hfill{$\Box$}
%Let $\bs{X}=(X_1,X_2...,X_n)$ be an i.i.d. sample from an absolutely continuous distribution function $F$.
%Let $U_n(\mu)$ with kernel $\Phi(\bs{X};\mu)$ be a $U$-statistic from the class $\mathcal{U}$, and let $\widehat{\mu}(\alpha)$, $0\leq \alpha\leq 1/2$, be the $\alpha$-trimmed sample mean.
%Then $\sqrt{n}U_n(\widehat{\mu}(\alpha)$ converges in distribution to a zero mean normal random variable with variance:

In the case of supremum-type statistics,
the following two theorems, analogous to the Theorem \ref{disperzijaModifikovane} and Theorem \ref{bocenjeno}, are used.

\begin{theorem}\label{raspodelaKS} Let $\bs{X}=(X_1,X_2...,X_n)$ be an i.i.d. sample from an absolutely continuous symmetric distribution, 
	with function $F$. 
	Let $\{U_n(\mu;t)\}$ be a non-degenerate family of U-statistics of order $m$ with kernel $\Phi(\cdot;t)$,
	that belong to the family $\mathcal{U}$; and let $\widehat{\mu}(\alpha)$, $0\leq \alpha\leq 0.5$, be the $\alpha$-trimmed sample mean 
	\eqref{alfasamplemean}.
	Then the family $\{U_n(\widehat{\mu}(\alpha);t)\}$ is also non-degenerate with the variance function

	\begin{align*}
	\sigma^2_{U,F}(\alpha;t)&=m^2\Bigg(\int_{-\infty}^{\infty}\varphi^2(x;t)f(x)dx+
	\bigg(\int_{-\infty}^{\infty}\varphi(x;t)f'(x)dx\bigg)^2\frac{2}{(1-2\alpha)^2}\\&\cdot\bigg(\int_{0}^{q_{1-\alpha}}x^2f(x)dx+\alpha(q_{1-\alpha})^2\bigg)
	+\frac{4}{1-2\alpha}\bigg(\int_{-\infty}^{\infty}\varphi(x;t)f'(x)dx\bigg)\\&\cdot\bigg(\int_{0}^{q_{1-\alpha}}\varphi(x;t)xf(x)dx+
	q_{1-\alpha}\int_{q_{1-\alpha}}^{\infty}\varphi(x;t)f(x)dx\bigg)\Bigg),
	\end{align*}
	for $0<\alpha<1/2$. In the case of boundary values of $\alpha$, we have
	
	\begin{align*}
	\sigma^2_{U,F}(0;t)&=m^2\Bigg(\int_{-\infty}^{\infty}\varphi^2(x;t)f(x)dx\\&+
	2\bigg(\int_{-\infty}^{\infty}\varphi(x;t)f'(x)dx\bigg)^2\bigg(\int_{0}^{\infty}x^2f(x)dx\bigg)
	\\&+4\bigg(\int_{-\infty}^{\infty}\varphi(x;t)f'(x)dx\bigg)\cdot\bigg(\int_{0}^{\infty}\varphi(x;t)xf(x)dx+
	\bigg)\Bigg),
	\end{align*}
	and
	\begin{align*}
	\sigma^2_{U,F}(1/2;t)&=m^2\Bigg(\int_{-\infty}^{\infty}\varphi^2(x;t)f(x)dx\\&+\frac{1}{4f^2(0)}
	\bigg(\int_{-\infty}^{\infty}\varphi(x;t)f'(x)dx\bigg)^2
	\\&+\frac{2}{f(0)}\bigg(\int_{-\infty}^{\infty}\varphi(x;t)f'(x)dx\bigg)\cdot\bigg(\int_{0}^{\infty}\varphi(x;t)f(x)dx\bigg)\Bigg).
	\end{align*}
	
	Moreover, $\sqrt{n}\sup|U_n(\widehat{\mu}(\alpha);t)|$ converges in distribution to the supremum of a certain centered Gaussian process.
	%Then
	%$\sqrt{n}(U_n(\mu),\widehat{\mu}(\alpha)-\mu)$ converges in distribution to bivariate normal $\mathcal{N}({\bf 0},\Sigma)$, where

\end{theorem}

\textbf{Proof.} The asymptotic  behaviour of  $\sqrt{n}U_n(\widehat{\mu}(\alpha);t)$ for a fixed $t$ is established in Theorem \ref{disperzijaModifikovane}.

From \cite[Theorem 2.8]{randles1982asymptotic} we have

\begin{align}\label{prf1}
U_n(\hat{\mu}(\alpha);t)\overset{d}{=}
U_n(\mu;t)+\eta_0(\hat{\mu}(\alpha);t)+R'_n,
\end{align}
where $\sqrt{n}R'_n\overset{p}{\to}0$ and $\eta_0(\mu;t)=EU_n(\mu;t)$.

Next, using the mean value theorem and the Bahadur representation \eqref{bahadurRepresentation}, we get
\begin{align}\label{prf2}
 \eta_0(\hat{\mu}(\alpha);t))&= \frac{\partial}{\partial\mu}\eta_0(\mu(\alpha);t)\frac1n\sum_{j=1}^n\psi_{\alpha,F}(X_j)+R''_n,
\end{align}

where $\sqrt{n}R''_n\overset{p}{\to}0$.

Combining \eqref{prf1} and \eqref{prf2} we get

\begin{align*}
\sqrt{n}U_n(\hat{\mu}(\alpha);t)&\overset{d}{=}
\sqrt{n}\Big(U_n(\mu;t)+\eta_0(\mu(\alpha);t)+\frac{\partial}{\partial\mu}\eta_0(\mu(\alpha);t)\frac1n\sum_{j=1}^n\psi_{\alpha,F}(X_j)\Big)+\sqrt{n}(R'_n+R''_n)\\
&=\sqrt{n}U^{*}_n(\mu(\alpha);t)+\sqrt{n}R_n.
\end{align*}

$U^{*}_n(\mu(\alpha);t)$ is asymptotically equivalent to a family of U-statistics with symmetrized kernel 
\begin{align*}
\Xi(\mathbf{X};\mu(\alpha),t)=\frac1m\frac{\partial}{\partial\mu}\eta_0(\mu(\alpha);t)\sum_{j=1}^m\psi_{\alpha,F}(X_j)+\Phi(\mathbf{X};\mu(\alpha),t).
\end{align*}

Using the result from \cite{silverman}, we have that $\sqrt{n}U^{*}_n$ converges in distribution to a zero mean Gaussian process. Then, since 
$\sqrt{n}R_n$ converges to zero in probability, using the Slutsky theorem for stochastic processes \cite[Theorem 7.15]{kosorok2008introduction},
we complete the proof.

\hfill{$\Box$}

\begin{theorem} Under the same assumptions as in Theorem \ref{raspodelaKS}, the limit in probability of the modified statistic  
	$\sup_{t}|U_n(\widehat{\mu}(\alpha);t)|$,
	under alternative  $g(x;\theta)\in \mathcal{G}$, is
	
	\begin{align*}
	b(\theta,\alpha)=m\sup\limits_{t}\Big|\int_{-\infty}^{\infty}\varphi(x;t)(h(x)+\mu'_{\theta}(0,\alpha)f'(x))dx\Big|\cdot\theta+o(\theta),
	\end{align*}
	
\end{theorem}
\textbf{Proof.}  The limit in probability of $U_n(\widehat{\mu}(\alpha);t)$ for a fixed $t$ is established in Theorem \ref{bocenjeno}. 
Denote $\eta(\mu;t)=E_{\theta}(U_n(\mu;t))$ and let $\eta(\widehat{\mu}(\alpha);t)$ be its estimator. 
From  \cite[Theorem 2.9]{iverson} we have that $U_n(\widehat{\mu}(\alpha);t)-\eta(\mu;t)=U_n(\mu;t)-\eta(\mu;t)+\eta(\widehat{\mu}(\alpha);t)-\eta(\mu;t)$ 
with probability one. Then
using  Glivenko-Cantelli theorem  for  U-statistics \cite{helmersjanssen}  we complete the proof. \hfill{$\Box$}

Finally, the following two theorems give us the Bahadur approximate slopes of the tests based on the Bonferoni measure and $\sqrt{b_1}$, respectively.

\begin{theorem}
	Let $(X_1,X_2...,X_n)$ be an i.i.d. sample with distribution function $G(x,\theta)\in \mathcal{G}$.  Then the Bahadur approximate slopes  of
	test statistics  CM, $\gamma$, and MGG are equal to
	
	\begin{align*}
	c(\theta)&=\frac{\Big(\int_{-\infty}^{\infty}xh(x)dx+\frac{H(0)}{f(0)}\Big)^2}{\sigma^2+\frac{1}{4f^2(0)}-\frac{\tau}{f(0)}}\cdot\theta^2+o(\theta^2),\theta\to 0,
	\end{align*}
	where $\sigma^2=\int_{-\infty}^{\infty}x^2f(x)dx$ and $\tau=2\int_{0}^{\infty}xf(x)dx$.

\end{theorem}

\textbf{Proof.}
We shall prove the theorem in the case of statistic CM. The other cases are completely analogous.  

Denote $D=\bar{X}-\hat{M}.$ Notice that $D$ is ancillary for the location parameter $\mu$. Hence, we may suppose that $\mu=0$. 
Using the Bahadur representation we have
\begin{equation*}
D=\frac{1}{n}\sum_{i=1}^n\Big(X_i-\frac{{\rm sgn}(X_i)}{2f(0)}\Big) + R_n,
\end{equation*}
where $\sqrt{n}R_n$ converges to zero in probability.
Using the central limit theorem   we have that the limiting distribution of $\sqrt{n}D$, when $n\to \infty$ is  normal with zero mean and the variance
\begin{equation}\label{disperzijaCM}
{\rm Var}\Big(X_1-\frac{{\rm sgn}(X_1)}{2f(0)}\Big)=\sigma^2+\frac{1}{4f^2(0)}-\frac{\tau}{f(0)}.
\end{equation}

Using the Slutsky theorem we obtain that the limiting distribution of $\sqrt{n}{\rm CM}$ is zero mean normal with the variance 
\begin{equation*}
\Big(\sigma^2+\frac{1}{4f^2(0)}-\frac{\tau}{f(0)}\Big)\sigma^{-1}.
\end{equation*}

Next, using the law of large numbers and the Slutsky theorem, we have that the limit in probability under a close alternative  $G(x,\theta)\in\mathcal{G}$ is
\begin{equation*}
b(\theta)=\frac{m(\theta)-\mu(\theta)}{\sigma(\theta)},
\end{equation*}
where $m(\theta), \mu(\theta)$ and $\sigma(\theta)$ are the mean, the median and the standard deviation, respectively.
Expanding $b(\theta)$ in the Maclaurin series,  and combining with \eqref{disperzijaCM} into \eqref{appslope},
we complete the proof. \hfill{$\Box$}

\begin{theorem}\label{teoremab1}
	Let $(X_1,X_2...,X_n)$ be an i.i.d. sample with distribution function $G(x,\theta)\in \mathcal{G}$.  Then the Bahadur approximate slope of
	test statistic  $\sqrt{b_1}$ is
	
	\begin{align*}
	c(\theta)&=\frac{\Big(\int_{-\infty}^{\infty} x^3h(x)dx-3{\sigma^2}\int_{-\infty}^{\infty} xh(x)dx\Big)^2}{m_6-6\sigma^2m_4+9\sigma^6}\cdot\theta^2+o(\theta^2),\theta\to 0,
	\end{align*}
	
	where $\sigma^2=\int_{-\infty}^{\infty}x^2f(x)dx$,
	and $m_j$ is the $j$th central moment of $F$.
\end{theorem}

The proof goes along the same lines as in the previous theorem, so we omit it here.
\section{Comparison of the Tests}

Since no test is distribution free, we need to choose the null variance in order to calculate the local approximate slope. Since we deal with the alternatives
close to symmetric, it is natural to choose the closest symmetric distribution for the null.

\subsection{Null and Alternative Hypotheses}

\begin{figure}[h!]
	\begin{center}
		
		\includegraphics[scale=0.6]{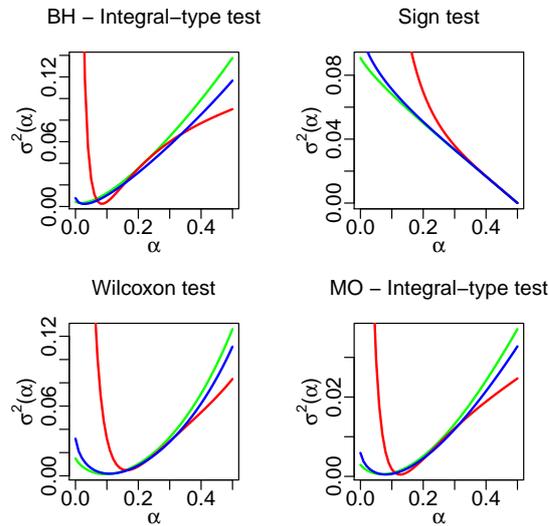}
		\label{fig: sigma}
		\caption{Limiting variances of integral test statistics --- green line -- normal;  blue line -- logistic; red line -- Cauchy;}
	\end{center}
\end{figure}

We consider the normal, the logistic and the Cauchy as  null distributions.
Using Theorem \ref{disperzijaModifikovane} we calculated the asymptotic variances of all our integral-type statistics, as well as the suprema
of variance functions of the supremum-type statistics. In Figure \ref{fig: sigma} we present the limiting variances of some integral-type statistics as
function of the trimming coefficient $\alpha$. It can be noticed that  for some values of $\alpha$ the variances are very close to each other. 
This "asymptotic quasi distribution freeness" might be of practical importance providing  an alternative to standard bootstrap procedures.

For each null distribution, we consider two types of close  alternatives from $\mathcal{G}$: % (see e.g. \cite{milosevicObradovicSimetrijaSPL}):

\begin{itemize}
	%\item a skew alternative in the sense of Azzalini \cite{azzalini} with the density
	%\begin{equation}\label{azz}
	%g(x;\theta)=2g(x;0)G(x\theta;0),\;\;\theta>0;
	%\end{equation}
	\item a skew alternative  in the sense of Fernandez-Steel \cite{fernandez}, with the density
	\begin{equation}\label{fernandez}g(x;\theta)=\frac{2}{1+\theta+\frac{1}{1+\theta}}\Big(g(\frac{x}{1+\theta};0)I\{x<0\}+g((1+\theta)x;0)I\{x\geq 0\}\Big)\end{equation}
	\item a contamination alternative with the density
	\begin{equation}\label{cont}
	g(x;\theta)=(1-\theta)g(x;0)+\theta g(x-1;0).
	\end{equation}
	
\end{itemize}
Another popular family of alternatives are the Azzalini skew alternatives (see \cite{azzalini2}). However, in the case of the skew-normal distribution,
all our test have zero efficiencies, while in the skew-Cauchy case, the  Bahadur efficiency is not defined. Hence, these alternatives are not  suitable 
for comparison, and we decided not to include them.

\subsection{Bahadur equivalence}

It turns out that some tests have identical Bahadur approximate slopes. With this in mind,
we say that two tests are Bahadur equivalent if their Bahadur local approximate slopes coincide. It is then sufficient to consider just one representative from each equivalence class for comparison purposes.

We have the following Bahadur equivalence classes:

\begin{itemize}
	\item $\rm BH^I \sim MO^I(1) \sim NA^I(2) \sim NA^I(3)$ for all $0\leq\alpha\leq 1/2$;
	\item $\rm BH^K \sim MO^K(1) \sim NA^K(2) \sim NA^K(3)$ for all $0\leq\alpha\leq 1/2$;
	\item $\rm CM \sim \gamma \sim MGG \sim S(\hat{\mu}(0))$;
	\item $\rm KS(\hat{\mu}(\alpha))\sim S(\hat{\mu}(\alpha))$  (up to a certain value of $\alpha$).
\end{itemize}
The first three equivalence classes can be easily obtained from the expressions for the corresponding Bahadur approximate  slopes.
For the fourth equivalence, notice that the term which is maximized in the KS  statistic, for $t=0$, is twice the absolute value of the S statistic. 
So, these tests are equivalent  whenever both the supremum of the asymptotic variance and the supremum of the limit in probability, are reached for $t=0$ 
(see Figure \ref{fig:  sigmaKS} as an example).
This is the case for small $\alpha$, from zero up to certain point that depends on the underlying null and alternative distributions.
%The first two follow directly from the definition of statistics or from equivalence of corresponding statistic without estimated parameters (see \cite{nikitinAhsanullah}).
% The equivalence $CM \sim S(\hat{\mu}(0))$ can be obtain directly from the formulas for asymptotic variances, while the other members of this class of equivalence are obtained using Slutsky theorem.

\begin{figure}[h!]
	\begin{center}
		\includegraphics[scale=0.6]{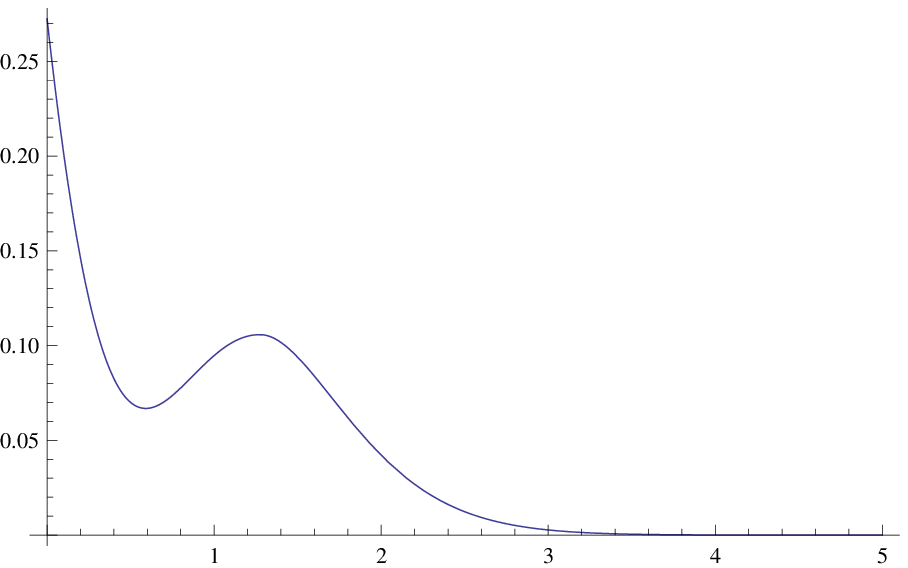}
		\includegraphics[scale=0.6]{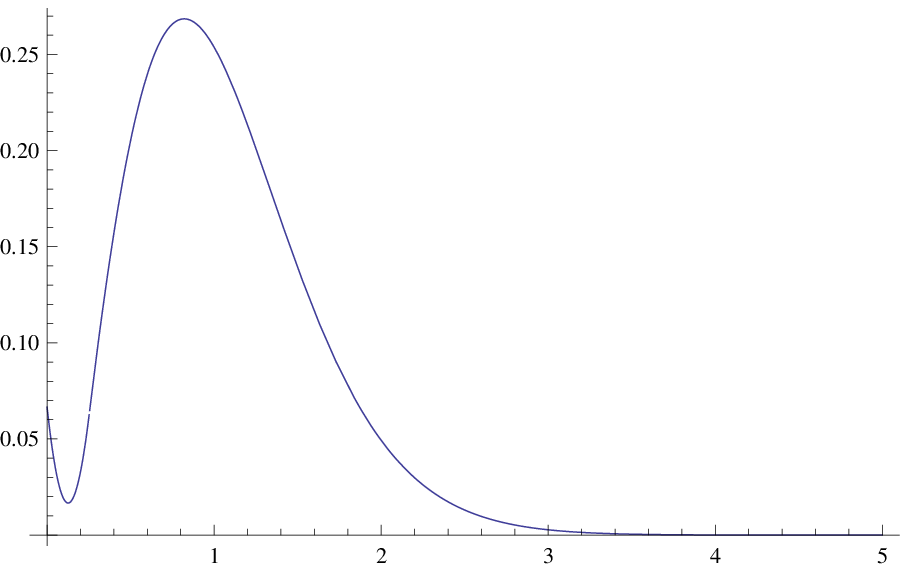}
		\label{fig: sigmaKS}
		\caption{Asymptotic variance functions $\sigma^2_{KS,F}(t)$ for $\alpha=0.1$ (left) and for  $\alpha=0.4$ (right) }
	\end{center}
\end{figure}

\subsection{Discussion}

Using Theorems \ref{disperzijaModifikovane}-\ref{teoremab1} we calculated the local approximate Bahadur slopes for all statistics, all null and all 
alternative distributions.

Taking into account the Bahadur equivalence, we choose the following tests: $\rm BH^I$ and $\rm BH^K$; $\rm NA^I(4)$ and $\rm NA^K(4)$ 
(denoted as $\rm NA-I$ and $\rm NA-K$); $\rm MO^I(2)$ and  $\rm MO^K(2)$ (denoted as $\rm MO-I$ and $\rm MO-K$); S; W; and KS. 
For the convenience in presentation, we display the Bahadur approximate indices graphically as functions of $\alpha$.  
We also present the indices of $\rm CM$ and $\sqrt{b_1}$ (denoted as $\rm b1$). Since they are not  functions of $\alpha$, we show
them as  horizontal lines.
\begin{figure}[h!]
	\begin{center}
		\includegraphics[scale=0.42]{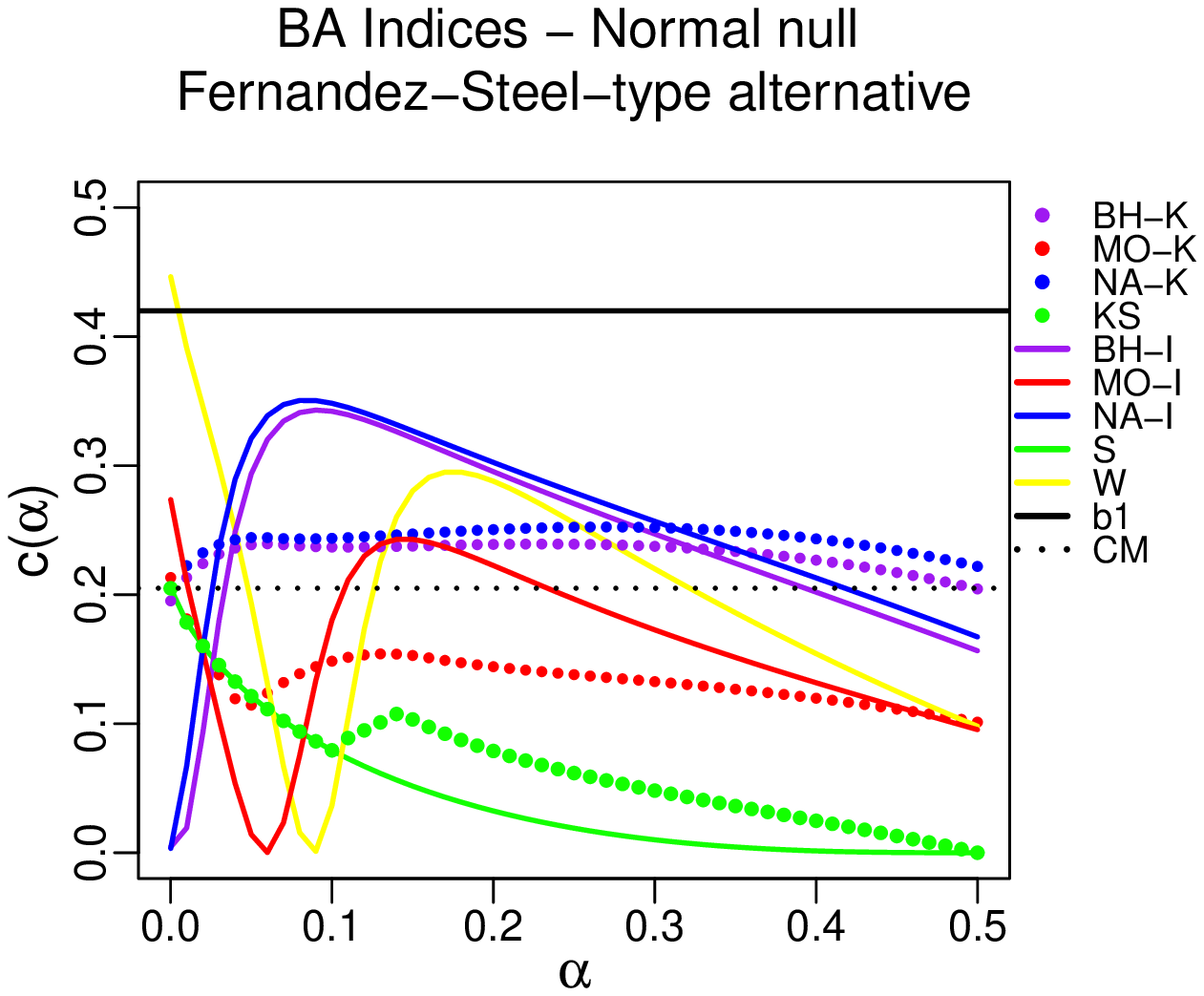}
		\includegraphics[scale=0.42]{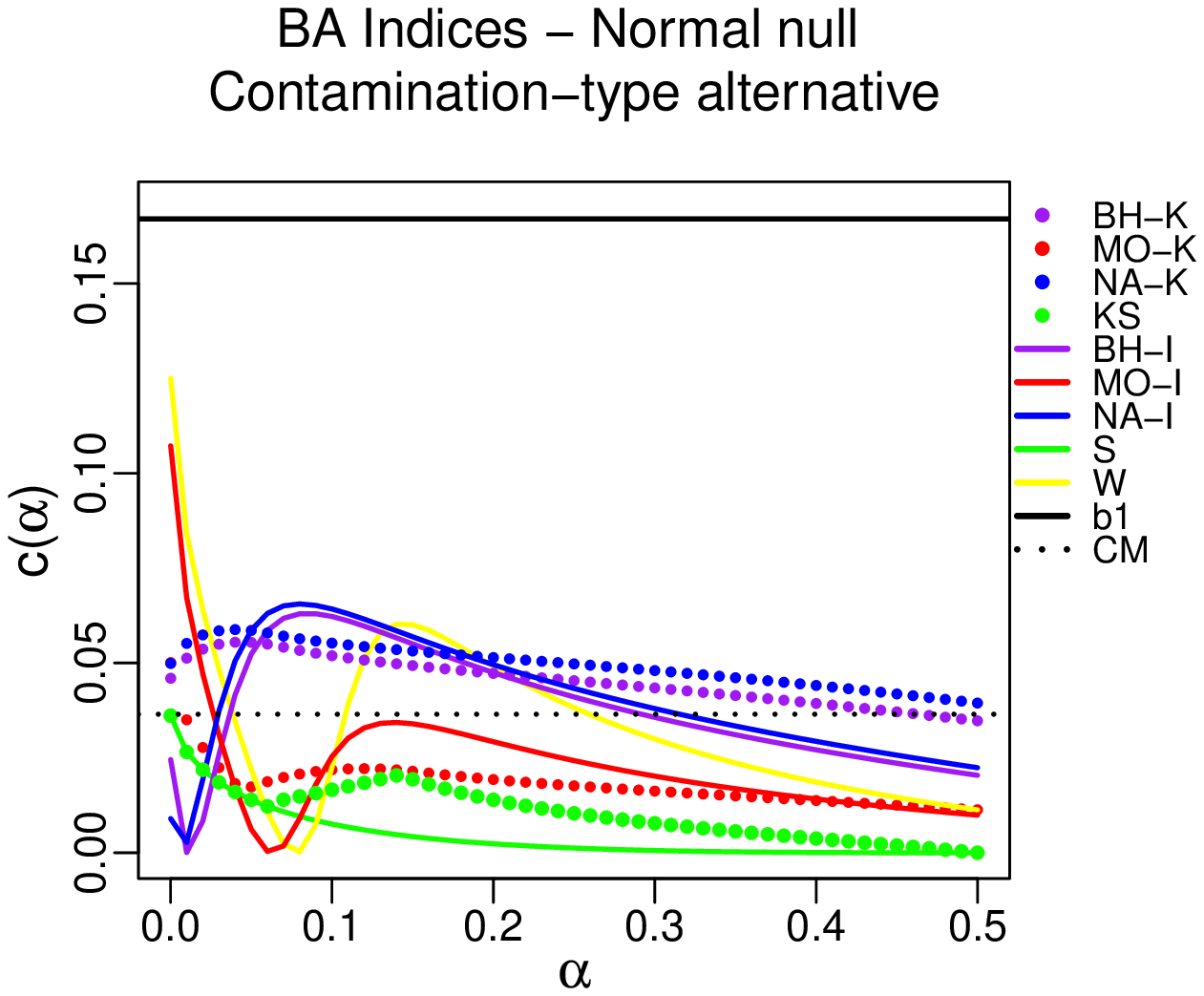}
		\label{fig: BAINormal}
		\caption{Comparison of Bahadur approximate indices -- normal distribution }
	\end{center}
\end{figure}

\begin{figure}[h!]
	\begin{center}
		\includegraphics[scale=0.42]{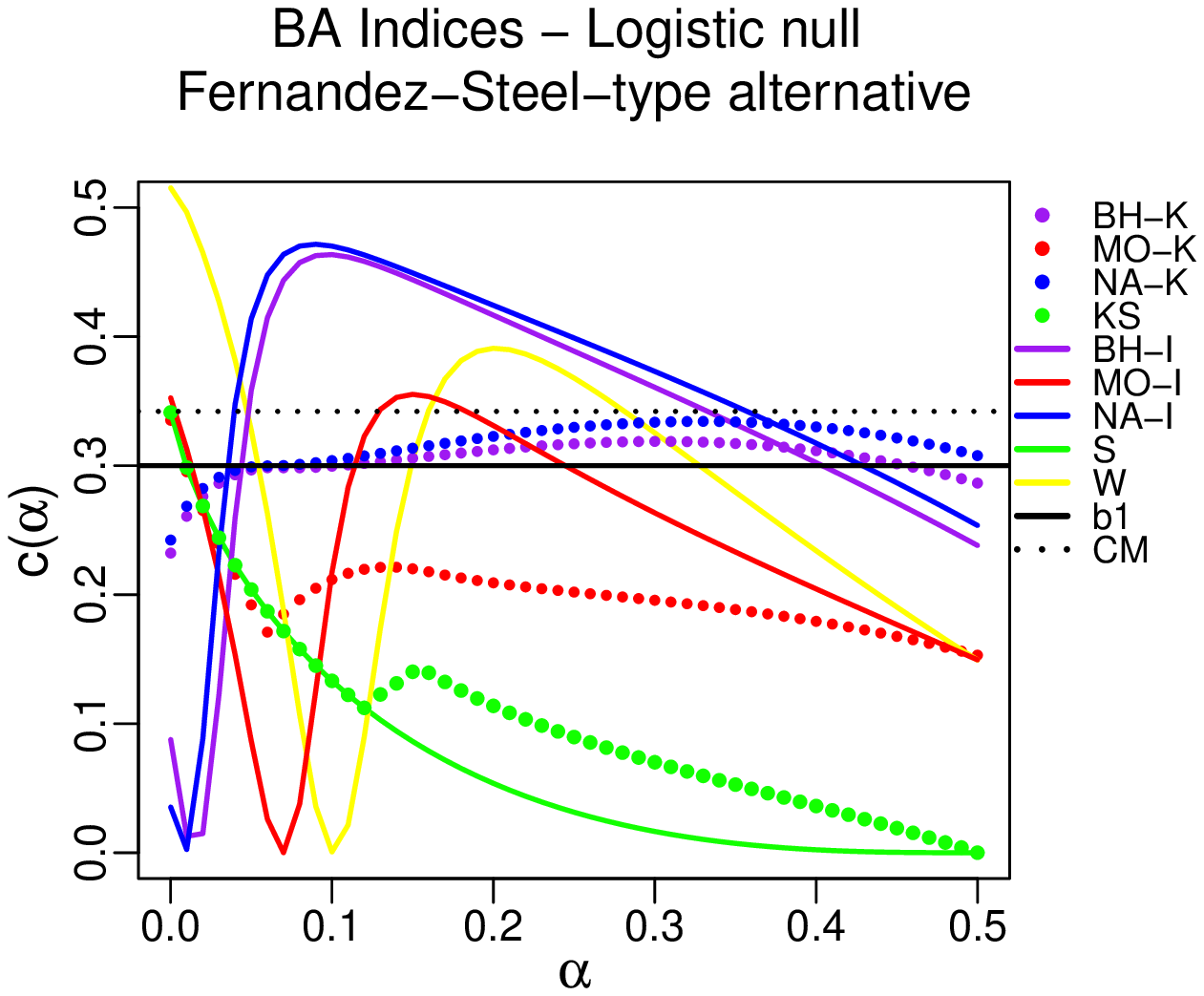}
		\includegraphics[scale=0.42]{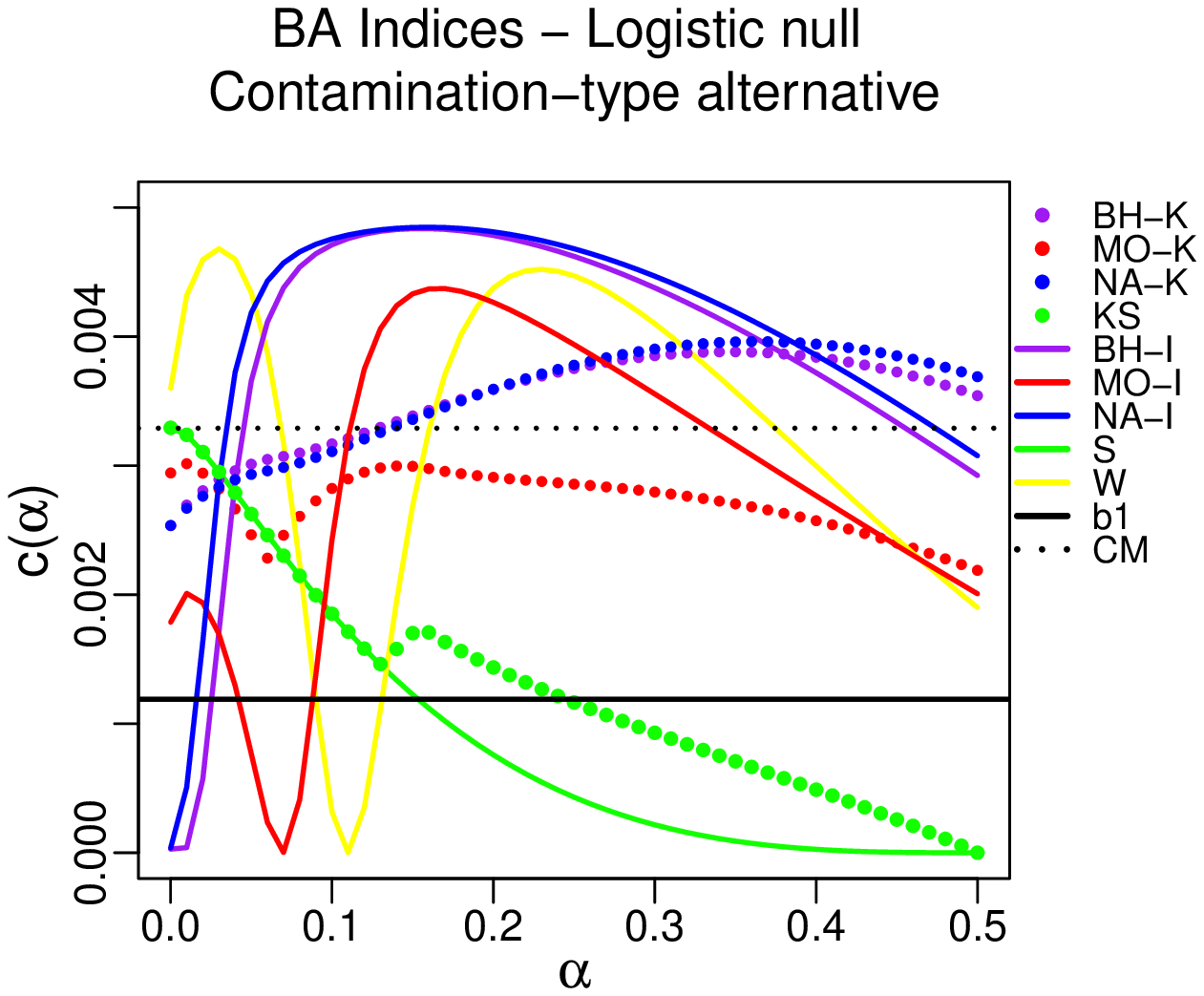}
		\label{fig: BAILogis}
		\caption{Comparison of Bahadur approximate indices -- logistic distribution}
	\end{center}
\end{figure}

\begin{figure}[h!]
	\begin{center}
		\includegraphics[scale=0.42]{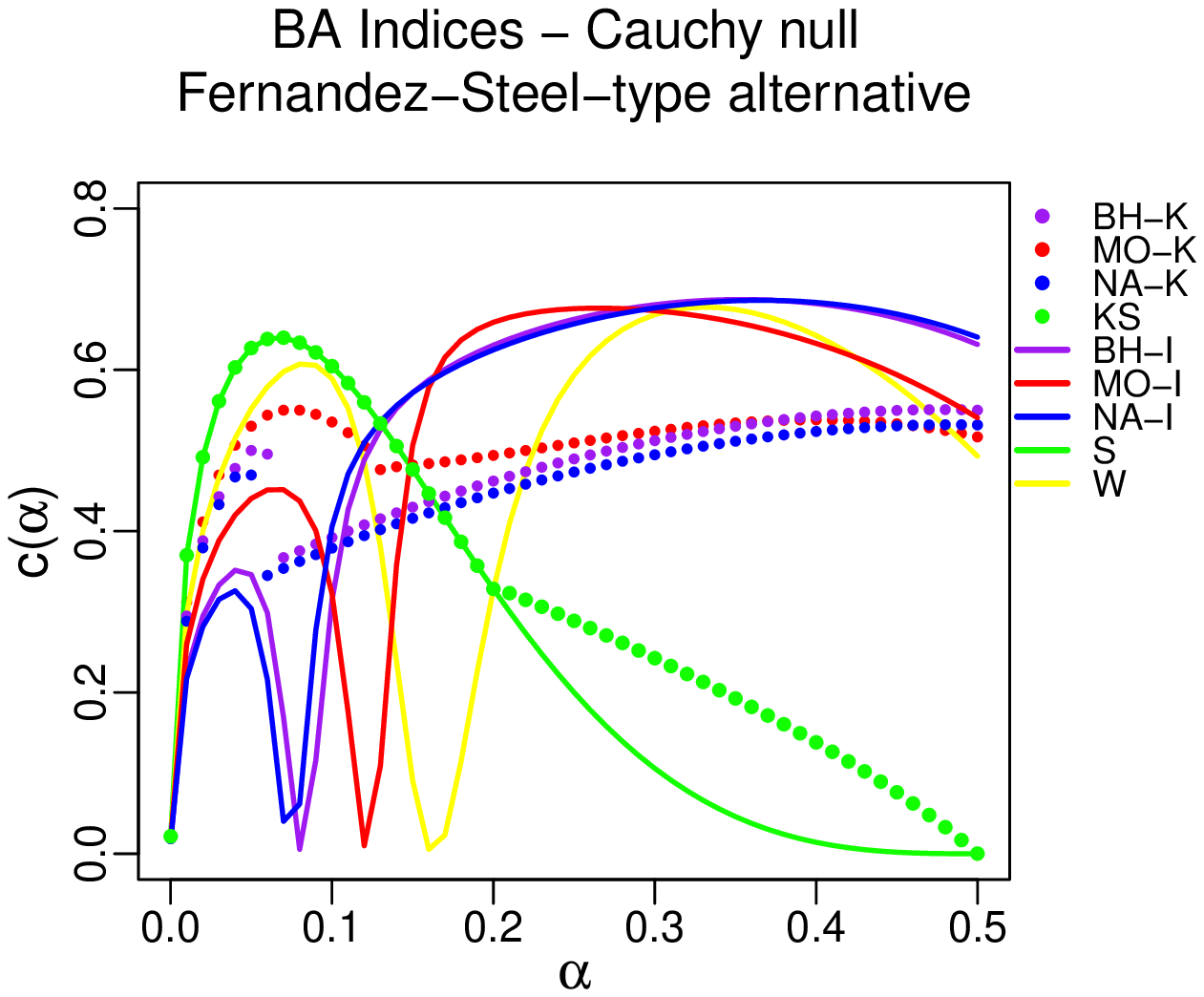}
		\includegraphics[scale=0.42]{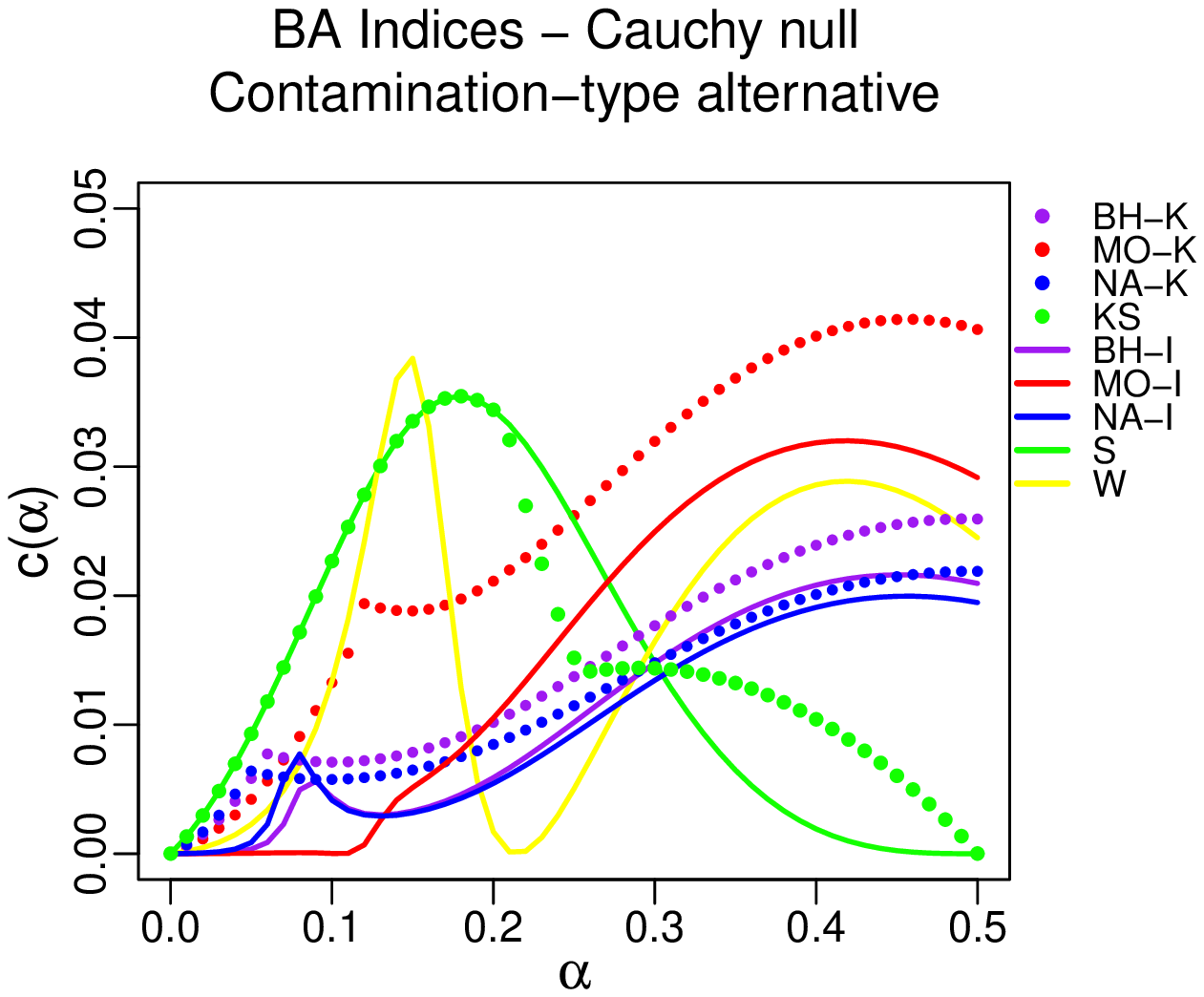}
		\label{fig: BAICauchy}
		\caption{Comparison of Bahadur approximate indices -- Cauchy distribution }
	\end{center}
\end{figure}

%From the Figures \ref{fig: BAINormal}

It is visible from all the figures that in the case of the integral-type statistics, the efficiencies vary significantly with $\alpha$. 
In particular, for all the tests exist a value of $\alpha$ for which they have zero efficiencies. On the other hand, the supremum-type tests are much 
less sensitive to the change of $\alpha$. The exception is the classical KS test which is by its definition inefficient for $\alpha=0.5$.

A natural way to compare tests, for a fixed null distribution, would be to compare the maximal values of their Bahadur  indices over  $\alpha\in [0,0.5]$. 
It can be noticed that, in most cases, the integral-type tests outperform the supremum-type ones. The only exception is the contamination alternative to the 
Cauchy distribution. This is in concordance with the previous results (see e.g. \cite{milosevicObradovicSimetrijaSPL,nikitinAhsanullah}).

In the case of normal distribution, the best of all tests are $\sqrt{b_1}$, and W for $\alpha=0$. In the case of the contamination alternative, 
$\rm MO^I(2)$ test for $\alpha=0$ is also competitive.
%This was, somewhat expected since the test was designed for it.
As far as the logistic distribution is concerned, $\rm NA^I(4)$ and $\rm BH^I$ are most efficient. In the case of the Cauchy null, the situation is different.
The tests CM and $\sqrt{b_1}$ are not applicable, and neither are the other  tests for $\alpha=0$. Also, the ``order'' of the tests is much different 
for the two considered alternatives. In case of the Fernandez-Steel  alternative, the best tests are  $\rm MO^I(2)$, $NA^I(4)$ and $\rm BH^I$, while in the 
case of the contamination alternative, $\rm MO^K(2)$ test is the most efficient.

As a conclusion, it is hard to recommend which test, and for which $\alpha$, is the best to use in general, when the underlying  distribution is completely 
unknown. The integral-type tests for small values of $\alpha$ could be the right choice, but they could also be calamities. In contrast, the supremum-type
tests with $\alpha$ close or equal to 0.5 are quite reasonable, and, most importantly, never a bad choice.
\section*{Acknowledgement}

We would like to thank the referees for their useful remarks.
This work was supported by the MNTRS, Serbia under  Grant No. 174012 (first author).



%\section*{References}
%\bibliographystyle{apalike}
\bibliographystyle{plain}
%\biboptions{authoryear,sort}
\bibliography{simetrija4}


%\begin{thebibliography}{00}


%\bibitem{}

%\end{thebibliography}
\end{document}